\documentclass[12pt]{iopart}

\usepackage{iopams}  

\usepackage{graphicx,xspace}
\usepackage{units}
\usepackage{hyperref}
\usepackage{epstopdf}

\usepackage{color,hyperref}
\definecolor{darkblue}{rgb}{0.0,0.0,1}
\hypersetup{colorlinks,breaklinks,
        linkcolor=darkblue,urlcolor=darkblue,
           anchorcolor=darkblue,citecolor=darkblue}

\newcommand{\ket}[1]{|#1\rangle}

\def\deg{$^\circ$C\xspace}

\begin{document}

\title[High-Fidelity Polarization Storage in a Gigahertz Bandwidth Quantum Memory]{High-Fidelity Polarization Storage in a Gigahertz Bandwidth Quantum Memory}

\author{D.~G.~England$^1$, P.~S.~Michelberger$^1$, T.~F.~M.~Champion$^1$, K.~F.~Reim$^{1, \ast}$, K.~C.~Lee$^1$, M.~R.~Sprague$^1$, X.-M.~Jin$^{1,2}$, N.~K.~Langford$^{1,\dagger}$, W.~S.~Kolthammer$^{1}$, J.~Nunn$^{1}$ and I.~A.~Walmsley$^1$}

\address{$^1$Clarendon Laboratory, Department of Physics, University of Oxford, Oxford, OX1 3PU, United Kingdom\\
$^2$ Centre for Quantum Technologies, National University of Singapore, 117543, Singapore\\
$^\ast$ Current address: Department of Physics, ETH Z\"{u}rich, CH-8093, Z\"{u}rich, Switzerland\\
$^\dag$ Current address: Department of Physics, Royal Holloway, University of London, Egham Hill, Egham TW20 0EX, United Kingdom}

\begin{abstract}


We demonstrate a dual-rail optical Raman memory inside a polarization interferometer; this enables us to store polarization-encoded information at GHz bandwidths in a room-temperature atomic ensemble. By performing full process tomography on the system we measure up to $97\pm 1\%$ process fidelity for the storage and retrieval process. At longer storage times, the process fidelity remains high, despite a loss of efficiency. The fidelity is $86\pm 4\%$ for \unit[1.5]{$\mu$s} storage time, which is 5,000 times the pulse duration. Hence high fidelity is combined with a large time-bandwidth product. This high performance, with an experimentally simple setup, demonstrates the suitability of the Raman memory for integration into large-scale quantum networks.

\end{abstract}

\maketitle

\section{Introduction}

Photons are well-established as carriers of quantum information and their transmission through high-bandwidth fibers or free-space opens the possibility of a global quantum network~\cite{Duan:2001p16886,Kimble:2008p17348}. To compensate for the effects of photon loss in a fiber network and the inherently probabilistic nature of quantum processes, it is necessary to map quantum information from a `flying' photonic qubit to a stationary one and back again, in a controllable manner. This is the essence of a quantum memory: it must faithfully store, and reproduce, the quantum state of a photonic qubit, including polarization. Key performance benchmarks of a quantum memory are high efficiency, long storage times and large bandwidths. Ultimately the clock-rate of a quantum information protocol will depend on the pulse durations used; a high bandwidth is required to store the temporally short pulses which lead to high processing rates. Also, in order to perform  operations while the qubit is in storage, the storage time of the memory must be several times larger than the pulse duration. For this reason, the time-bandwidth product (TBP) is an important metric for a quantum memory.  Note that the TBP, which represents the number of clock cycles spanned by the storage time, is distinct from the multimode capacity which represents the number of temporal or spectral bins that can be simultaneously stored. A large TBP is critical for synchronisation tasks, while large multimode capacity is advantageous for some protocols that invoke multiplexing. 

Quantum memories have been demonstrated in ultracold atoms~\cite{Liu:2001p13466,Zhao:2008p17809,Zhao:2008p17808}, cryogenically cooled solids~\cite{Longdell:2005p25,Chaneliere:2010p17807,Afzelius:2010p17806} and single atoms in high-finesse optical cavities~\cite{Maitre:1997p13519}. However, if quantum memories are to be used as part of a global quantum network, they will eventually have to operate in remote, unmanned locations, so the apparatus must be simple and robust. A promising candidate for such robust operation is a room-temperature atomic ensemble. Storage times of several milliseconds have been achieved using electromagnetically induced transparency (EIT) in simple vapor cells~\cite{Julsgaard:2004p16892}. The gradient-echo memory (GEM) technique has also been used to great effect in atomic gases, achieving up to 87\% readout efficiency utilizing a switched magnetic field gradient~\cite{Hosseini:2009p17793,Hosseini:2011p17792}. Despite these long storage times, and high efficiencies, the bandwidth of atomic memories is often limited by the narrow linewidth of the atomic transitions, hence precluding a large TBP. However this limitation can be overcome by using a controlled read-in and read-out mechanism based on an off-resonant Raman interaction, which has been proposed~\cite{Nunn:2007p13183} and demonstrated~\cite{Reim:2010p13180} by this group. 

\begin{figure}[h!]
\centering
\includegraphics[width=0.8\textwidth]{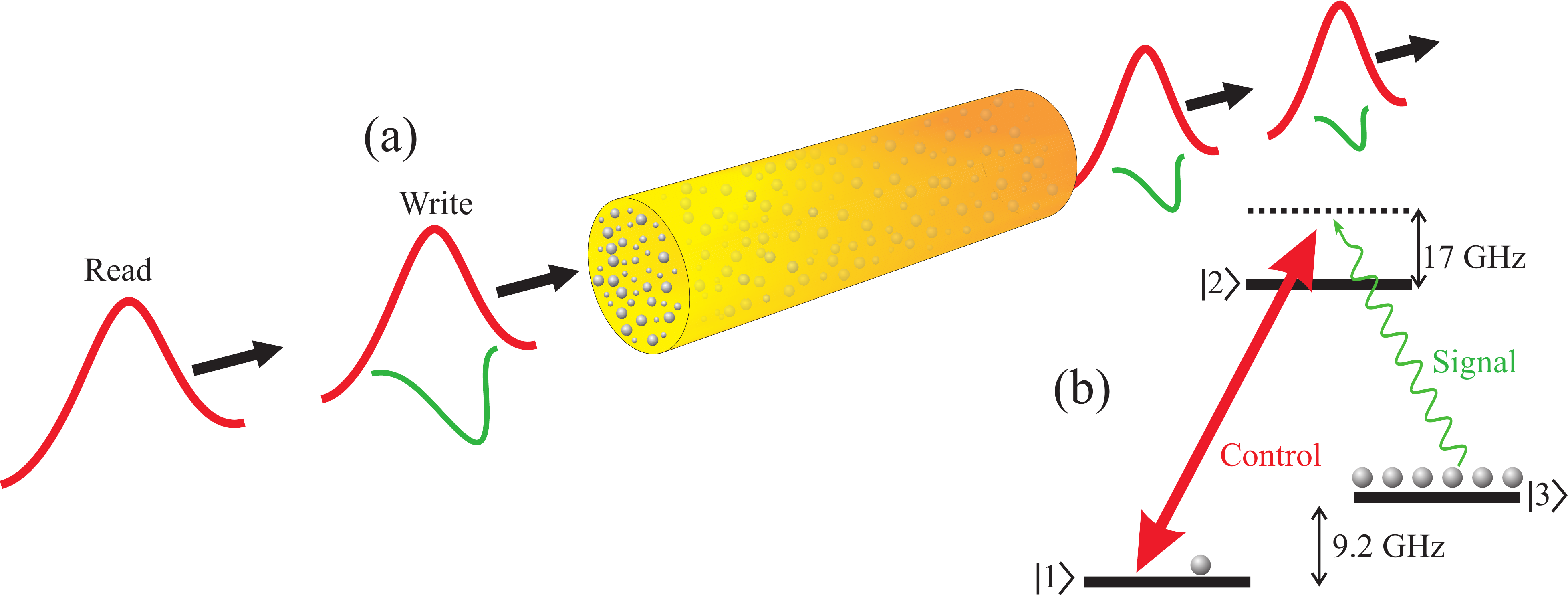}
\caption[]{{\bf (a)} The Raman quantum memory scheme: A weak signal pulse (thin green) is mapped into a collective excitation of an atomic ensemble by a strong orthogonally polarized control pulse (thick red). Upon further application of a second control pulse, the signal is read out of the memory due to imperfect memory efficiency some signal light is no t stored; this is transmitted by the memory. {\bf (b)} Level diagram for the Raman memory operating on the Cesium $D_2$ line. The signal is stored via an off-resonant Raman transition in a $\Lambda$-level system in atomic Cesium vapor. The ground states $6S_{1/2}, F=3$ (denoted by $\ket{3}$) and $6S_{1/2}, F=4$ ($\ket{1}$) are split by \unit[9.2]{GHz} and the detuning from the excited state, $6P_{3/2}$ ($\ket{2}$), is typically around \unit[17]{GHz}. Prior to the memory operation, the atoms are prepared in $\ket{1}$ by optical pumping.}
\label{fig:Mem}
\end{figure}

The Raman memory protocol is based on a two-photon off-resonant process in an atomic $\Lambda$-level system, with a weak signal and strong control pulse, which maps the electric field of the signal pulse onto a collective excitation in an atomic ensemble known as a {\em spin wave}, as shown in figure \ref{fig:Mem}. Here the strong control pulse produces a virtual excited state, with a linewidth determined only by the bandwidth of the control pulse. Because the control and signal fields must address the two ground-states separately, the bandwidth of this memory is limited, in practice, by the ground state splitting. Based on this interaction, we have implemented a memory in room-temperature Cesium (Cs) vapor~\cite{Reim:2010p13180,Reim:2011p13181}, which is capable of storing pulses of \unit[300]{ps} duration, corresponding to a \unit[1.5]{GHz} bandwidth with a maximum efficiency of 30\%. The maximum storage time, currently limited by residual magnetic fields, is approximately \unit[2]{$\mu$s}, which is $10^4$ times longer then the pulse duration. These parameters yield the highest TBP of any quantum memory so far. By attenuating the signal field such that it contained, on average, 1.6 photons per pulse, we demonstrated memory operation on the single-photon level. The unconditional noise floor on these measurements was \unit[0.25]{photons/pulse}, indicating the functionality of this system in the quantum regime. The origins of this noise are expected to be spontaneous Raman scattering and four-wave mixing~\cite{Reim:2011p13181}, the former could be removed by improved frequency filtering, but the latter is intrinsic to the vapor cell memory. Raman quantum memory represents a robust and reliable option for integration in large-scale quantum networks. Its high bandwidth, coupled with an unprecedented TBP, and the abililty to store single photons makes it ideally suited to synchronizing probabilistic quantum events, for example in entanglement swapping or enhancing multi-photon rates. 

To interface effectively with future quantum networks, a memory must also be capable of storing the quantum information encoded in the incoming photons.  In fiber-based networks, the polarization degree of freedom is particularly useful, because it allows a single-photon qubit to be transmitted in a single spatial and temporal mode.  Furthermore, because photons generally interact only weakly with their environment, polarization-encoded information can be transmitted over long distances without decoherence --- for example, radiation from the Big Bang is still partially polarized~\cite{Kogut:2003p17131}.  This is a critical requirement for the feasibility of large-scale networks.  The ability to store polarization information with high fidelity is, therefore, a key benchmark for a useful quantum memory.

Raman memory may operate in a multimode configuration, but in its simplest form it is effectively single-mode~\cite{Nunn:2008p17074}; hence it cannot store an arbitrary polarization state. However, this problem can be addressed by building a dual-rail memory inside a polarization interferometer with one arm storing the vertical component of the polarization state and the other the horizontal. In this way the polarization state of the input light can be perfectly stored in the two ensembles. This has been successfully demonstrated using the EIT technique in ultracold atoms~\cite{Jin:2010p17638,Chou:2007p15771,Choi:2008p17756} and a warm atomic ensemble~\cite{Cho:2010p16987}. Unlike the EIT memory, the novel off-resonant nature of the Raman memory allows the storage of high-bandwidth pulses. Furthermore, a low unconditional noise floor due to the suppressed collisional fluorescence facilitates single-photon level operation at room-temperature~\cite{Reim:2011p13181}, which is not possible in many schemes, and technically challenging in others. These advantages motivate the investigation of polarization storage in the Raman memory. In this paper we demonstrate that the Raman memory can store polarization-encoded information using the dual-rail procedure. We perform state tomography on the dual-rail Raman memory to characterize its capability to store polarization. We observe a process fidelity of up to $97\pm1\%$ at \unit[12.5]{ns} storage time. This fidelity remains high for longer storage times, yielding $86\pm4\%$ after \unit[1.5]{$\mu$s} --- 5000 times longer than the pulse duration. Hence we demonstrate high fidelity polarization storage at large bandwidths with unprecedented time-bandwidth products.

\section{Dual-rail memory \label{sec_TwoModeMemory}}

The experimental implementation of the Raman memory in warm Cesium gas is discussed elsewhere~\cite{Reim:2010p13180}, so is only briefly described here. The master laser for this experiment is a Spectra Physics Tsunami which produces pulses of \unit[300]{ps} duration at a \unit[80]{MHz} repetition rate. The laser is tuned close to the Cesium $D_2$ line at \unit[852]{nm}. A Pockels cell picks two pulses from this 80 MHz pulse train with a variable delay, where the first pulse defines the storage and the second the retrieval time window. From these picked pulses, the orthogonally polarized signal and control pulses are derived by a polarizing beam splitter. The signal is subsequently shifted by the Cs hyperfine ground state splitting of \unit[9.2]{GHz}, using an electro-optic modulator (EOM), to obtain two-photon Raman resonance. The EOM is time gated, such that only pulses in the storage time window are frequency modulated, yielding the presence of signal field only in the storage time bin. Subsequently, the signal and control arms are re-overlapped spatially and temporally on a PBS and are focussed into the Cs vapor cell, which is heated to 67\deg using resistive heating coils. The Raman interaction only couples orthogonal polarizations far from resonance, hence the signal and control fields remain orthogonal. In the memory output mode, this orthogonality enables extinction of the strong control field. Frequency filters are used to further extinguish the control field before detecting the signal on a photodiode. Prior to the memory procedure, the atomic ensemble is prepared in the memory ground state ($6 \rm{S}_{1/2}, \rm{F=4}$) by optical pumping with a diode laser. The optical pumping laser is orthogonally polarized to the signal and is in a counter-propagating geometry. The cell is shielded from stray magnetic fields with several layers of $\mu$-metal. By fitting the storage time to our model of magnetic dephasing, we estimate the stray magnetic field to be on the order of \unit[0.1]{Gauss}~\cite{Reim:2011p13181}, which is consistent with residual magnetization generated by our heating coils.

\begin{figure}[h!]
\centering
\includegraphics[width=0.8\textwidth]{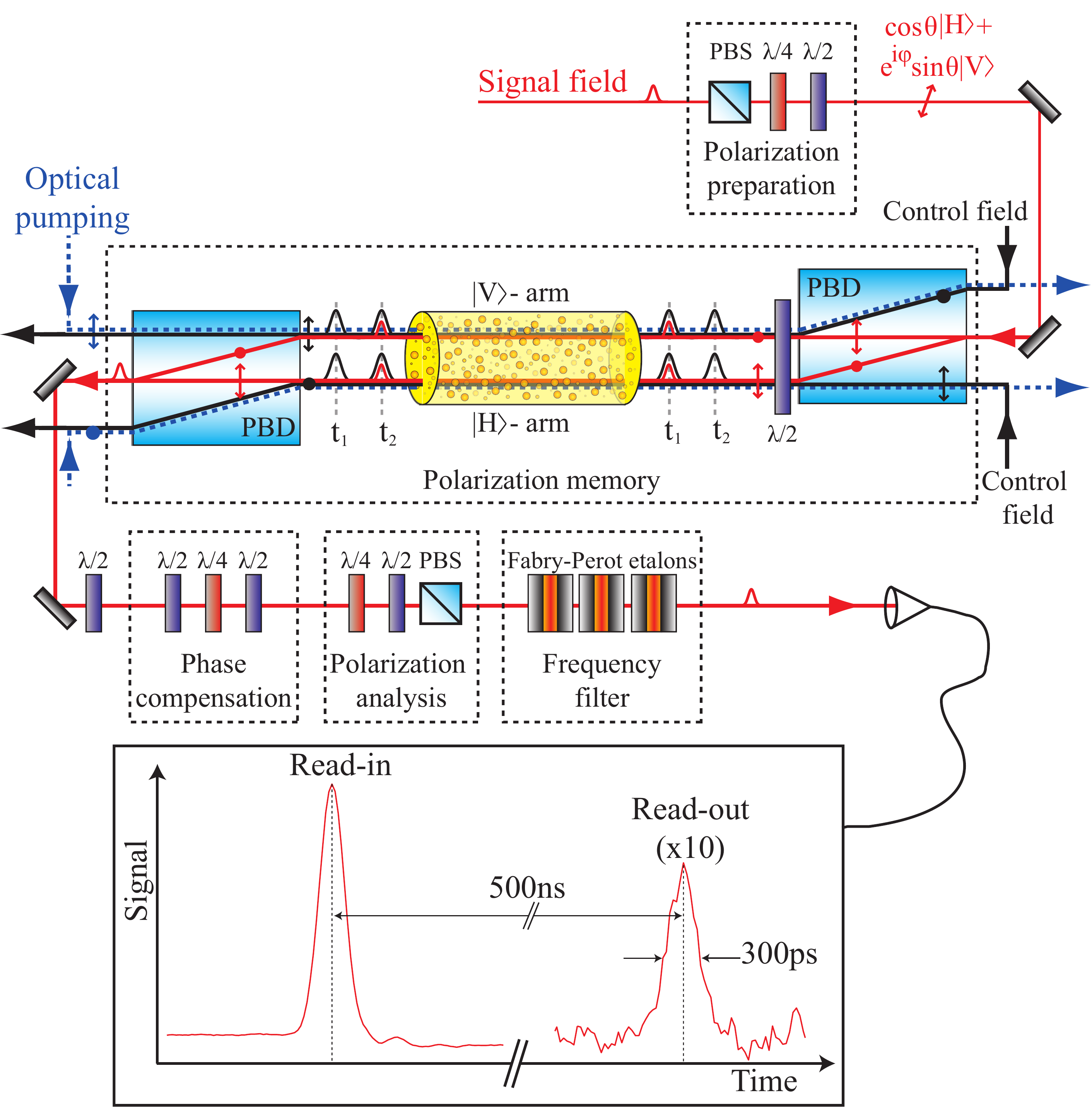}
\caption[]{Layout of the two-memory interferometer. The signal field (red) is prepared in an arbitrary polarization state, $\cos{\theta}\ket{H} + e^{i\phi}\sin{\theta}\ket{V}$, which is split into two arms in a polarization interferometer using a pair of polarizing beam displacers (PBD). The control field (black) and optical pumping laser (blue dashed) are introduced along each arm with the orthogonal polarization, hence enter and exit the PBDs at different ports. The signal is read in at time $t_1$, and out at time $t_2$, by a strong control pulse. Following the memory, the signal field polarization is analyzed with a polarizer and a pair of calibrated waveplates before Fabry--Perot etalons are used to extinguish stray control field light. Additional phase picked up in the interferometer is compensated by a pair of quarter-wave plates set to $\pm$45$\rm{^\circ}$ sandwiching a half-wave plate.  The signal and control field preparation, as well as the focussing lenses are omitted for clarity, for details refer to reference~[14]. The inset shows a typical memory signal, in this instance the storage time is \unit[500]{ns}, the read-out signal is magnified by a factor of 10 for clarity.}
\label{fig:PBD_mem}
\end{figure}

Since, in this regime, the Raman memory is single-mode, polarization-encoded information cannot be stored in a single atomic ensemble. Instead, we construct a passively stable polarization interferometer, employing two polarizing beam displacers (PBD)~\cite{Cho:2010p16987,Obrien:2003p17856}, with the Cs cell positioned inside the interferometer (see fig. \ref{fig:PBD_mem}). The PBDs split the signal pulses into their constituent horizontal (H) and vertical (V) components inside the interferometer with subsequent recombination at the interferometer output. The orthogonally polarized optical pumping laser and control pulses are overlapped with the signal field in the interferometer but are spatially separated outside of it. The phase accumulated by the signal due to unequal path lengths between the H and V arms is compensated behind the interferometer output~\cite{Bhandari:1988p14296}. In this way, we create a two-mode memory by accessing two non-overlapping atomic ensembles in the same Cs cell, with one mode storing horizontally (H) and the other vertically (V) polarized light. By balancing the efficiencies of these two memories, which prevents an artificial rotation of the output polarization, we can accurately store polarization information.

\section{Quantum process tomography}
\label{sec_QPT} 

In order to benchmark the polarization storage capability of the memory, we use quantum process tomography (QPT)~\cite{Chuang:1997p2455,OBrien:2004p14369,Lobino:2008p17804,RahimiKeshari:2011p17805}.  A quantum process is any physical process, unitary or non-unitary, which takes a physical input state $\rho_{\rm{in}}$ and produces a physical output state $\rho_{\rm{out}}$.  In the quantum process formalism~\cite{Nielsen:2003p14361}, any such process can be written as
\begin{equation}
\label{eq:QPT1}
\rho_{\rm out} = \sum_{i,j}\chi_{ij}\Gamma_i\rho_{\rm in}\Gamma_j,
\label{eq_QPTformula}
\end{equation}
where $\chi_{ij}$ is known as the {\em process matrix} and contains the complete information about the dynamics of the system, and the $\Gamma_{i,j}$ are a complete set of orthonormal basis matrices for the density matrix.  QPT is a technique for estimating an unknown quantum process by preparing a range of different input states and making measurements on the output state.  For our polarization qubit, we prepare and measure six polarization states: $\{ \ket{H}, \ket{V}, \ket{D} = \frac{1}{\sqrt{2}}(\ket{H} + \ket{V}), \ket{A} = \frac{1}{\sqrt{2}}(\ket{H} - \ket{V}), \ket{R} = \frac{1}{\sqrt{2}}(\ket{H} -i \ket{V}), \ket{L} = \frac{1}{\sqrt{2}}(\ket{H} + i\ket{V}) \}$.  The 36 resulting input-output measurement settings provide a reliable basis set with which to fully characterize the storage process for the qubit system~\cite{deBurgh:2008p052122, Adamson:2010p14531}.  Here, we reconstruct the measured process matrix using maximum likelihood estimation~\cite{OBrien:2004p14369} (for a detailed recipe, see ~\cite{LangfordNK2007phd}).  Each measurement was repeated 500 times to determine the measurement statistics, which were in turn used to determine error bars via Monte Carlo simulation.

To characterize our qubit memory, we measure the process with the memory switched ``on'', by analyzing the retrieved signal field with the control field present, and ``off'', by analyzing the transmitted signal field with the control field blocked.  We then quantify the memory performance by calculating the process fidelity between the two resulting process matrices, defined by $F = {\rm tr}\left[\sqrt{\sqrt{\chi_{\rm on}}\chi_{\rm off}\sqrt{\chi_{\rm on}}}  \right]^2$~\cite{Gilchrist:2005p062310}, a measure of the similarity of two different quantum processes which in our case is synonymous with the memory's ability to preserve polarization encoded information.  Hence, for an ideal Raman memory, the ``on'' process is identical to the ``off'' process, which should just be the identity process, corresponding to $F=1$. Fidelities of less than 1 imply that the state has been altered by the process.

\section{Results}
The experiment is performed using classical weak coherent states containing on the order of 1000-10,000 photons per pulse. However, these results will also hold for truly quantum single-photon inputs because the counting statistics of single photons passing through a linear optical system always follow the classical behavior~\cite{Loudon:2004p14489}. An obvious example of this being the interference visibility of light attenuated below the single-photon-level~\cite{Taylor:1909p14511}. In order to run the experiment in the single-photon regime, the interferometer would have to be modified to include small waveplates in each arm to compensate for the birefringence of the cell windows. Currently, this birefringence causes a small rotation of the control field polarization leading to leakage through the polarization filtering; this added control field noise precludes single-photon level measurements. In addition, the long counting times required to build up single-photon statistics require stability of the interferometer on long time scales of several hours, which would require active stabilization of the interferometer.

\begin{figure}[h!]
\centering
\includegraphics[width=0.7\textwidth]{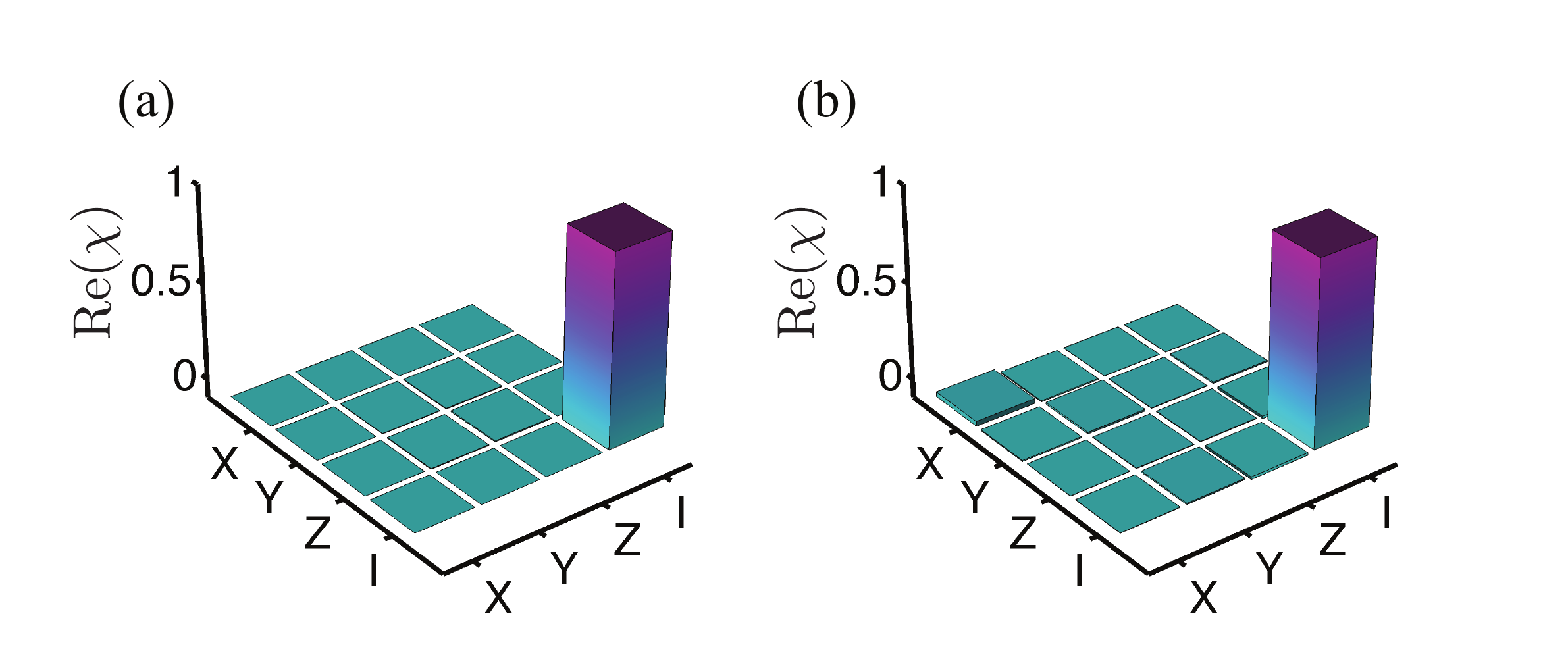}
\caption[]{{\bf (a)} The process matrix, $\chi_{\rm off}$, as measured with the control field blocked. This evaluates the performance of the polarization interferometer. As can be seen, only the identity transformation appears in the process, proving the stability of the interferometer during the measurement. {\bf (b)} The process matrix of the retrieved pulse, $\chi_{\rm on}$, after \unit[750]{ns} storage time. The fidelity of the memory process is calculated, by comparison of these two matrices (see sec. \ref{sec_QPT}), to be 85$\pm$4\%.}
\label{fig:QPT}
\end{figure}

To assess the coherence of the polarization storage, full process tomography of the memory was performed at a range of storage times and the fidelity is obtained from the reconstructed process matrices $\chi_{ij}$, as described in sec. \ref{sec_QPT}. By running the experiment in its ``off'' state, with the control field blocked, we also obtain the process matrix of the interferometer. Figure \ref{fig:QPT} shows typical reconstructed process matrices for the input (control blocked) and retrieved pulse for an exemplary storage time of \unit[750]{ns}. The input process matrix, $\chi_{\rm off}$, can be seen to consist mainly of the identity, $I$, demonstrating that the interferometer is stable throughout the measurement. The retrieved matrix, $\chi_{\rm on}$, is also dominated by the identity showing that the memory replicates the polarization state faithfully. In this case, the process fidelity was calculated to be 85$\pm$4\%.

\begin{figure}[h!]
\centering
\includegraphics[width=0.6\textwidth]{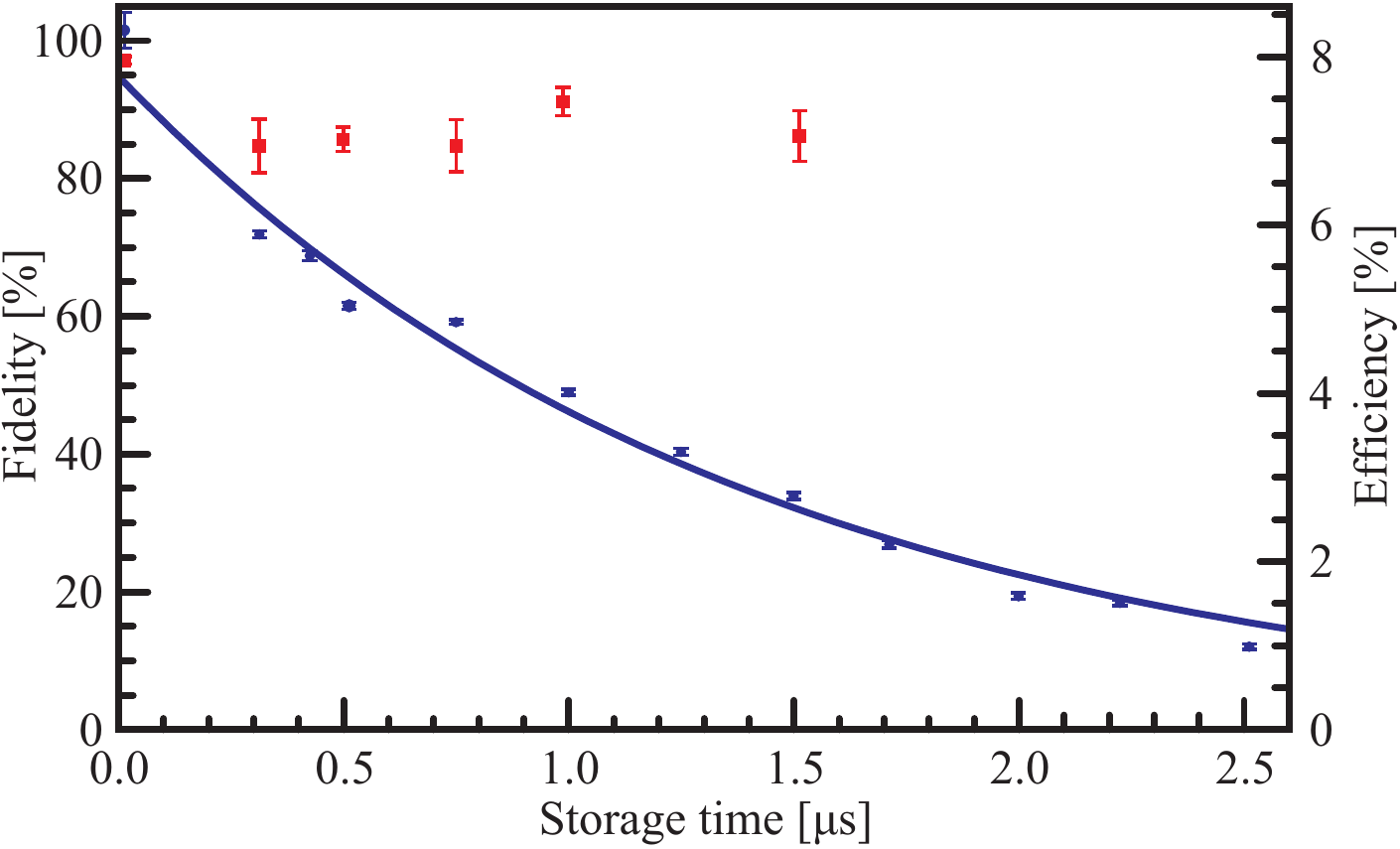}
\caption[]{The process fidelity of the memory (red squares) plotted alongside the efficiency of the memory (blue dots) with increasing storage time. The fidelity remains high even as memory efficiency decreases, implying that the fidelity of stored polarization information is insensitive to loss. The efficiency is lower here than has previously been reported~\cite{Reim:2011p13181} as the control field power is shared between both arms of the polarization interferometer.}
\label{fig:Visibility}
\end{figure}

The individual values for the process fidelity are plotted, alongside the memory efficiency, as a function of storage time in Figure \ref{fig:Visibility}. This shows that the fidelity was highest for \unit[12.5]{ns} storage ($97\pm1\%$), but remained constantly above 84\% for storage times of up to \unit[1.5]{$\mu$s}, beyond which the retrieved signal became too weak for a meaningful reconstruction of the process matrix due to the decreasing memory efficiency (see fig. \ref{fig:Visibility}).  Notably the fidelity is approximately constant as a function of storage time. Thus it does not degrade despite the decreasing efficiency of the memory, showing that no polarization coherence, and hence no information, is destroyed by memory losses. This is an important feature if this memory is to be integrated in future quantum networks, illustrating the high quality and the robustness of this memory protocol for polarization information with respect to decoherence.

From previous studies, we expect the limiting factor in the storage time to be stray magnetic fields~\cite{Reim:2011p13181} as a non-zero magnetic field causes dephasing of the spin wave inside the atomic ensemble and therefore a loss of efficiency. However, a consistent fidelity seems to suggest that the spinwave remains coherent, despite the decrease in efficiency. One way to resolve this apparent discrepancy is to consider the effect of the control pulse. The spin-wave is created, and read-out, by pulses of the same polarization. For this reason the read-out pulse selects only the component of the spin wave which remains coherent with the read-in. This causes the read-out signal to have the same polarization as the input signal, hence the polarization state, and thereby the process fidelity, are maintained.

The off-resonant nature of the Raman memory means that, ordinarily, the ensemble is transparent to the signal field. An advantage of this feature is that unstored signal photons are simply transmitted by the memory (as shown in fig. 1) and can be used in subsequent experiments. To confirm that these transmitted photons are not affected by the action of the control pulse, the process fidelity of the unstored photons was also calculated. Typically, a fidelity of over 99.5\% was measured indicating that the transmitted photons are unaffected by the storage process. 

\section{Conclusion}
In conclusion, we have demonstrated the storage and on-demand retrieval of polarization-encoded information in a room-temperature Raman quantum memory with high fidelity at GHz bandwidths. The high time-bandwidth product and rugged design of this memory make it a promising candidate for integration in scalable quantum networks. The polarization basis represents a reliable and robust option for the transmission of photonic quantum information. Thus the preservation of polarization during storage and retrieval is of paramount importance for a quantum memory. In this paper we have performed process tomography on a dual-rail Raman quantum memory, demonstrating storage of the polarization of a weak coherent state with up to $97\pm1\%$ process fidelity. The fidelity remains above 84\% for up to \unit[1.5]{$\mu$s} storage time which is around 5000 times longer than the pulse duration, so this high-fidelity storage is coupled with a record time-bandwidth product. Furthermore, the fidelity does not decrease with increasing storage time, despite losses in memory efficiency, showing that the fidelity of the information remaining in storage is insensitive to loss. 

The off-resonant operation of the Raman memory suppresses collision-induced fluorescence, making single-photon storage and retrieval possible with low noise. This has already been demonstrated in a single-mode Raman memory. Although the polarization memory was implemented here with weak coherent states, only technical difficulties preclude the storage of single photons in the dual-rail memory. Hence this result represents a key step towards the storage of true single-photon polarization qubits. 

\section*{Acknowledgements}
This work was supported by EPSRC (project number EP/C51933/01) and the European Community's Seventh Framework Programme FP7/2007-2013 under grant agreement n$^\circ$ 248095 for project Q-ESSENCE, and the Royal Society. KFR was supported by the Marie-Curie Initial Training Network (ITN) EMALI and PSM was supported by the ITN FASTQUAST. XMJ acknowledges support from the Centre for Quantum Technologies at the National University of Singapore. MRS was supported by a Clarendon scholarship.

\section*{References}

\bibliographystyle{unsrt}

\bibliography{QPT_Bib}

\end{document}